%% file: InterspeechTemplate.tex
\documentclass[a4paper]{article}

\usepackage{INTERSPEECH2020}

\usepackage{multirow}
\usepackage{cite}
\usepackage[colorlinks]{hyperref}
\usepackage[table,xcdraw]{xcolor}
\usepackage{amsmath,graphicx}
\usepackage{subfig}
\usepackage[capitalise]{cleveref}
\usepackage[printonlyused]{acronym} 
\input{acrodef}

\usepackage{colortbl}
\usepackage{hhline}
\usepackage{enumitem}
\newcommand{\ts}{\textsubscript}
\newcolumntype{C}[1]{>{\centering\let\newline\\\arraybackslash\hspace{0pt}}m{#1}}

\title{Ensemble of Discriminators for Domain Adaptation \\ in Multiple Sound Source 2D Localization}

\name{
Guillaume Le Moing$^{1,2}$
, Don Joven Agravante$^{1}$, Tadanobu Inoue$^{1}$, Jayakorn Vongkulbhisal$^{1}$, \\
Asim Munawar$^{1}$, Ryuki Tachibana$^{1}$, Phongtharin Vinayavekhin$^{1}$
}

\address{
  $^{1}$IBM Research, Tokyo, Japan\\
  $^{2}$MINES ParisTech - PSL Research University, Paris, France
}

\email{guillaume.le\_moing@mines-paristech.fr, pvmilk@ibm.com}

\begin{document}

\maketitle

\begin{abstract}
This paper introduces an ensemble of discriminators that improves the accuracy of a domain adaptation technique for the localization of multiple sound sources.
Recently, deep neural networks have led to promising results for this task, yet they require a large amount of labeled data for training.
Recording and labeling such datasets is very costly, especially because data needs to be diverse enough to cover different acoustic conditions.
In this paper, we leverage acoustic simulators to inexpensively generate labeled training samples.
However, models trained on synthetic data tend to perform poorly with real-world recordings due to the domain mismatch.
For this, we explore two domain adaptation methods using adversarial learning for sound source localization which use labeled synthetic data and unlabeled real data.
We propose a novel ensemble approach that combines discriminators applied at different feature levels of the localization model.
Experiments show that our ensemble discrimination method significantly improves the localization performance without requiring any label from the real data.

\end{abstract}


\section{Introduction}
\label{sec:intro}

\ac{SSL} aims at estimating a pose/location of sound sources.
With an increasing popularity in installing smart speakers in home environments, source location provides additional knowledge that could enable a variety of applications such as monitoring human activities in daily life~\cite{Vacher2011}, speech enhancement~\cite{jeyasingh2018} and human-robot interaction~\cite{jour:csl:argentieri2015}.
\ac{SSL} is an active research topic for which various signal processing methods have been proposed~\cite{jour:csl:argentieri2015, jour:wcmc:cobos2017}.
These data-independent methods work well under strict assumptions~\cite{jour:wcmc:cobos2017}, e.g.\ high signal-to-noise ratio, known number of sources, low reverberation, etc.
Such ideal conditions hardly hold true in real-world applications and usually require special treatments~\cite{conf:interspeech:netsch2014, jour:signal:ma2006,jour:tasl:Alexandridis2018, conf:interspeech:guo2016}.
Recently, data-driven approaches and in particular deep learning have outperformed classical signal processing methods for various audio tasks
\cite{jour:ispl:salamon2017,tech:dcase:Inoue2018} including
\ac{SSL}~\cite{conf:icassp:pertila2017,
conf:eusipco:adavanne2018,
conf:interspeech:he2018,
conf:icra:he2018,
conf:icassp:takeda2016,
jour:sensors:diaz2018,
conf:ica:pujol2019,
jour:aslp:ma2017}.


Multiple network architectures have been proposed to localize sound sources.
An advantage of these methods, apart from their ability to adapt to challenging 
acoustic conditions and microphone configurations, is that they can be trained 
to solve multiple tasks at the same time like simultaneous localization and 
classification of sounds~\cite{conf:interspeech:he2018}.
However, a significant downside is that they require lots of training data, 
which is expensive to gather and label~\cite{conf:icassp:He2019, conf:mmsp:lemoing2019}
Acoustic simulators are an appealing solution as they can abundantly generate 
high-quality labeled datasets.
However, models trained on synthetic data as a source domain can suffer from a 
critical drop in performance when exposed to real-world data as the target 
domain. This is due to acoustic conditions which are outside the distribution of 
the synthetic training dataset~\cite{jour:jrm:yalta2017,conf:icassp:carlo2019},
thus resulting in a \textit{domain shift}~\cite{jour:springer:ben-david2010}.

Recently, there have been several works about domain adaptation for \ac{SSL}. 
For example,~\cite{conf:icassp:Takeda2017, conf:icassp:Takeda2018} proposed 
unsupervised methods using entropy minimization of the localization output.
However, such methods are not suitable to our problem because entropy 
minimization encourages the prediction of only a single 
source whereas we must cater to multiple outputs.
In this context, \cite{conf:icassp:He2019} has proposed two adaptation 
methods compatible for multiple \ac{SSL}; (\textit{m}$_1$) a weakly supervised 
method in which the number of sources is provided for real-world data; (\textit{m}$_2$) an 
unsupervised method based on \ac{DANN}~\cite{jour:jmlr:Ganin2015} which intends 
to align latent feature distributions for synthetic and real-world domains by 
adding a discriminative model at a certain feature level of the localization 
model.
They reported that \textit{m}$_1$ increased the localization performance whereas \textit{m}$_2$ did 
not yield significant improvements.
However, adversarial methods such as~\cite{jour:jmlr:Ganin2015} are popular 
outside \ac{SSL}. For example,~\cite{jour:coor:Vu2018} proposes an adversarial 
domain adaptation method for the semantic segmentation problem in 
computer vision.
Moreover, similar approaches have been successfully applied to other audio tasks such as Acoustic Scene
Classification (ASC)~\cite{conf:icassp:wei2020} and Sound Event Detection (SED)~\cite{workshop:dcase:Gharib2018}.
Since it is not clear whether adversarial methods are suitable for \ac{SSL}, we also present extensive experiments with such methods.

\begin{figure*}[ht]
    \centering
    \includegraphics[width=\textwidth]{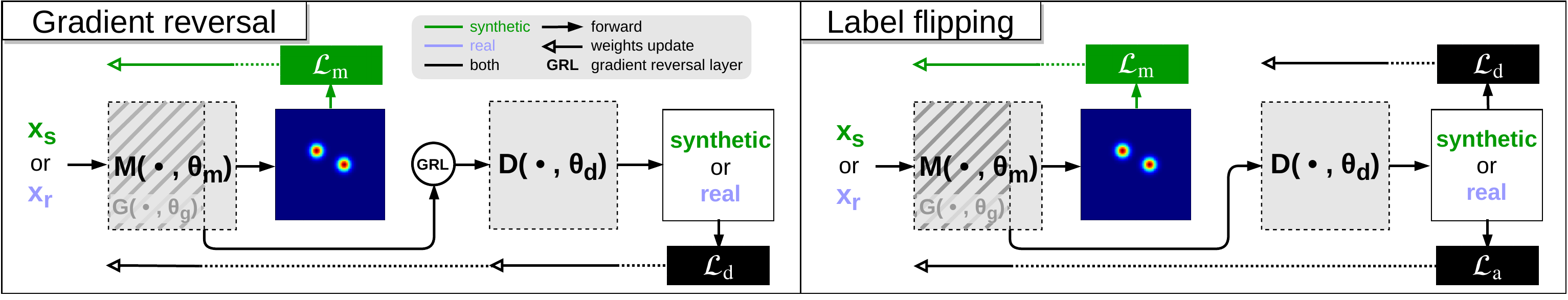}
    \caption{Outline of the domain adaptation framework for the two studied methods with intermediate discrimination.}
    \label{fig:label_flipping_gradient_reversal}
\end{figure*}

In this paper, we tackle the \textit{lack-of-data} problem for 2D multiple \ac{SSL} by leveraging simulation to cheaply generate a large amount of labeled data.
We then address the domain shift issue by training the localization model in such a way that the knowledge is transferable to the real-world.
The main contributions of this work are:
(a) We explore two domain adaptation methods based on adversarial learning: gradient reversal and label flipping. We analyse the difference of the two methods and show that they work for multiple \ac{SSL}.
(b) We propose a novel ensemble discrimination method for domain adaptation in \ac{SSL}. It consists of placing a discriminator at different levels of the localization model: the intermediate and output space.
(c) We evaluate our proposed method extensively on data captured from a real environment.
For each method, we ensure the consistency of the results by performing the experiments on different random seeds and reporting their mean and variance.
Through our experiments, we show that our ensemble discrimination method outperforms the standard adversarial learning approaches and can perform similarly to localization models that were trained in a supervised fashion on labeled real-world data.
To the best of our knowledge, this is the first successful attempt in tackling domain adaptation for multiple \ac{SSL} that does not require any label from the real-world data.

\section{Domain Adaptation for Multiple Sound Source 2D Localization}
\label{sec:adaptation-method}

Supervised learning relies on the assumption that data is independently and 
identically distributed. So training a deep learning model on synthetic data 
and testing without further 
training on real-world data usually results in a performance drop due to the
distributions being different, that is, the simulator is not perfect.
However, improving the simulation model is often too difficult. A more 
practical approach is randomizing the 
simulator parameters~\cite{jour:iros:tobin2017}. This \textit{expands} the 
source domain in the hopes of covering the 
target domain.
In acoustic simulators, parameters such as noise 
and reverberation can be randomized to get this effect.
However, this can still be insufficient for 
more difficult tasks~\cite{jour:iros:tobin2017}.
Assuming we can no longer improve the simulator and the model performance on 
real data is still insufficient, we need to use domain adaptation 
techniques~\cite{jour:springer:ben-david2010}.
Such methods consists in making the model aware of the domain mismatch during 
the training process.
This paper focuses on these methods, and in particular those that use 
adversarial learning.

To start, we formally describe the base task of learning to predict the 
location of multiple simultaneous sound sources in a 2D horizontal plane.
First, we extract spectral features $\mathcal{X}$ from the sound recorded by a 
set of microphones. Then, a localization model $M$, with parameters $\theta_m$, 
is trained to map spectral features to a localization heatmap $\mathcal{Y}$: a 
discretized grid with Gaussian distributions centered at the source locations. 
Training the model amounts to solving:
\begin{equation}
	\min_{\theta_m}\sum_{x\in\mathcal{X}}\mathcal{L}_\text{m}(M(x), y)
	,
    \label{eq:loc-obj}
\end{equation}
where $y \in \mathcal{Y}$ and the \ac{MSE} is used as the localization model loss 
($\mathcal{L}_\text{m}$). Such a framework was introduced in 
\cite{conf:mmsp:lemoing2019}, to which readers are referred for further 
details. 

We wish to augment (\ref{eq:loc-obj}) so as to perform well on real-world data 
$\mathcal{X}_r$ while only 
having labels $\mathcal{Y}_s$ for the synthetic data $\mathcal{X}_s$.
Without labels $\mathcal{Y}_r$, (\ref{eq:loc-obj}) cannot be used directly to learn
the mapping from real-world sound features to location cues.
This is an \textit{Unsupervised Domain Adaptation} problem. To solve 
this, we use the adversarial learning approach to make 
features generalizable and indistinguishable between the \textit{synthetic} and 
\textit{real} domains.
A discriminative model $D$, with parameters $\theta_d$, is plugged at a 
given layer of the localization neural network model.
Namely, we cut the neural network at a given stage, and feature maps outputted
from this layer are fed not only to the second part of the localization model but
also to the discriminator model.
We denote the subpart of the localisation model up to this layer as $G$, 
with parameters $\theta_g$ (a subset of $\theta_m$).
\cref{fig:label_flipping_gradient_reversal} shows this framework.
Here, $M$ is always trained with (\ref{eq:loc-obj}), using only 
synthetic labeled data.
$D$ is trained to assign domain class labels ($1$ for synthetic domain and $0$ 
for real-world domain) by using a \ac{BCE} loss ($\mathcal{L}_\text{d}$).
Meanwhile, $G$ tries to generate features that cannot be distinguished by $D$.
This can be formalized as a minimax objective~\cite{conf:nips:goodfellow2014} 
which is solved to jointly train $G$ and $D$:
\begin{equation}
  \small
  \min_{\theta_d}\max_{\theta_g}
  \sum_{x_s\in\mathcal{X}_s}
  \mathcal{L}_\text{d}(D(G({x_s})),1)
  +
  \sum_{x_r\in\mathcal{X}_r}
  \mathcal{L}_\text{d}(D(G({x_r})), 0)
  .
  \label{eq:minimax-obj}
\end{equation}
One cannot straightforwardly implement (\ref{eq:minimax-obj}), because 
backpropagation algorithms are meant to deal with cost-function minimization. 
There are two ways to solve this in practice.

\textbf{Gradient reversal} was introduced to train 
\ac{DANN}~\cite{jour:jmlr:Ganin2015} where it is called the \ac{GRL}. 
It is an unweighted layer, placed between $G$ and $D$. It behaves as the 
identity function during forward pass and negates the gradients during 
backward pass. This allows optimizing the weights $\theta_g$ by changing the 
max to a min.

\textbf{Label flipping} is another method that is commonly used by the 
adversarial learning community. It consists in decomposing the minimax 
objective (\ref{eq:minimax-obj}), 
into two minimization problems. Both are expressed as:
\begin{equation}
	\min_{\theta}
	\sum_{x_s\in\mathcal{X}_s}
	\mathcal{L}_\text{d}(D(G({x_s})), \alpha)
	+
	\sum_{x_r\in\mathcal{X}_r}
	\mathcal{L}_\text{d}(D(G({x_r})), \beta)
	,
    \label{eq:dis-obj}
\end{equation}
where the equation changes depending on the setting for $(\theta,\alpha,\beta)$.
For optimizing $\theta_d$, we set it to $(\theta_d,1,0)$ and refer to this as
(\ref{eq:dis-obj}a).
To optimize $\theta_g$, we flip the 
labels by using the setting $(\theta_g,0,1)$ and refer to it 
as (\ref{eq:dis-obj}b).
Furthermore, to help illustrate this difference in 
\cref{fig:label_flipping_gradient_reversal}, we also use $\mathcal{L}_\text{a}$ 
(adversarial loss) instead of $\mathcal{L}_\text{d}$ when using 
(\ref{eq:dis-obj}b).

Although both \ac{GRL} and label flipping are methods that intend to solve 
(\ref{eq:minimax-obj}), there are important differences.
Firstly, gradient reversal only requires one forward 
/ backward pass in $D$ to solve (\ref{eq:minimax-obj}). Whereas label 
flipping requires two passes, that is, one 
for each objective (\ref{eq:dis-obj}a), (\ref{eq:dis-obj}b). 
Secondly, the gradient computation differs in the magnitude at each update 
step. 
For label flipping, the weights update for $G$ and $D$ follows the same dynamic 
with respect to their objective so that the update is larger the farther away 
it is from the optimum.
In contrast, the dynamic is inverted for gradient reversal for updates 
on $\theta_g$. This results in smaller updates farther 
away from the optimum, hence slowing down convergence.
This can cause $D$ to converge faster than $G$ which may result in $D$ not being able to 
provide sufficient gradients to improve $G$~\cite{conf:nips:goodfellow2014}.
Although we present this basic analysis, the global effect on the learning 
dynamics of the complete \ac{SSL} model is unclear. Furthermore, stable 
adversarial learning is still an active research topic. Due to this, we evaluate 
both methods by extensive empirical results for our 
task of \ac{SSL}.






\textbf{Ensemble of Discriminators:} Discrimination level refers to the layer of $M$ at which we do 
adversarial learning. Although this can be continuously moved, we 
opt to take only the two extremes. First is \ac{dint}, where we place 
the discriminator, $D$, right after the \textit{encoder} of the \ac{SSL} model 
in~\cite{conf:mmsp:lemoing2019}. In this case, the submodel $G$ is the encoder.
We do not go further into the 
encoder to allow enough \textit{capacity} for $G$ to learn the domain 
independent features.
Second is \ac{dout}, where we placed $D$ after the output layer 
such that $G$ is all of $M$.
A prerequisite for the success of this method is that the distribution of sound
sources (number and 2D location of sources) in $\mathcal{Y}_s$ and $\mathcal{Y}_r$
are very similar.
If not, the discriminator can learn the dissimilarities and the generator will 
impair its localization predictions in order to satisfy the minimax objective 
(\ref{eq:minimax-obj}).
Constraining the output is common for domain adaptation in \ac{SSL} with, for 
example, entropy methods~\cite{conf:icassp:Takeda2017, conf:icassp:Takeda2018} 
or 
weak supervision~\cite{conf:icassp:He2019}.
However, such methods can degrade performance by 
boosting incorrect low confidence predictions~\cite{conf:icassp:He2019} 
resulting in a higher risk of false positives.
Using \ac{dout} has this same concern.
Lastly, we proposed a type of novel \textit{ensemble} method by using both 
\ac{dint} and \ac{dout}. For this, the two discriminators are 
trained independently but there is only one $M$ trained by both adversarial 
losses together with 
$\mathcal{L}_\text{m}$. Our intuition is that doing so would improve $M$ in different ways leading to better results.

\section{Experiments}


\subsection{Dataset}

Our data-collection setup was completely described 
in~\cite{conf:mmsp:lemoing2019} and we recall the main points here.
Experiments are done in a $6 \text{m} \times 6 \text{m}$ area with two 
microphone arrays. 
Music clips from the classical-funk genres are used for recording 
training, validation and testing data 
splits. An additional testing dataset from a jazz genre was collected to 
further verify the generalizability of the model.
One or two sources can be active simultaneously.
We set the minimum distance between simultaneous sources to be $2 \text{m}$ to ease 
the 
evaluation process.



For synthetic data, we used Pyroomacoustics~\cite{conf:icassp:scheibler2018}, 
where we simulated a model similar to the real room.
We generated $10^6$ samples with labels in an anechoic chamber with low noise 
setting as a training dataset, \textbf{TrainS}.
We also synthesized another dataset with the same amount of data, 
\textbf{TrainS+}, 
by randomizing the distance to the walls around the area from $0$m up to $10$m 
and the SNR in a log-scale from $0.1$ up to $1$.



For testing and comparing, we also need real-world data.
To ease labeling,
we used the augmentation method
described in~\cite{conf:mmsp:lemoing2019},
wherein we only captured one active 
source and then used
sound superposition to generate samples with multiple sources.
In each $1 \text{m} \times 1 \text{m}$ cell of the area, we randomly selected and recorded $7, 1, 1$ 
positions for training, validation and 
testing 
data, respectively.
Each $0.16$s audio sample was generated by randomly choosing or mixing one or 
more segments of the audio clips.
\textbf{TrainR}, \textbf{ValidR}, \textbf{TestRc} and \textbf{TestRj} are the 
real-world datasets for training, validating and testing respectively.
All real-world datasets were generated from the classical-funk music clips, 
except for TestRj which is based on jazz.
We generated $10^6$ samples for training sets and $5\times10^3$ for validation 
and testing sets.

\begin{figure}%
    \centering
    \subfloat[TestRc]{
        \includegraphics[width=7.9cm]{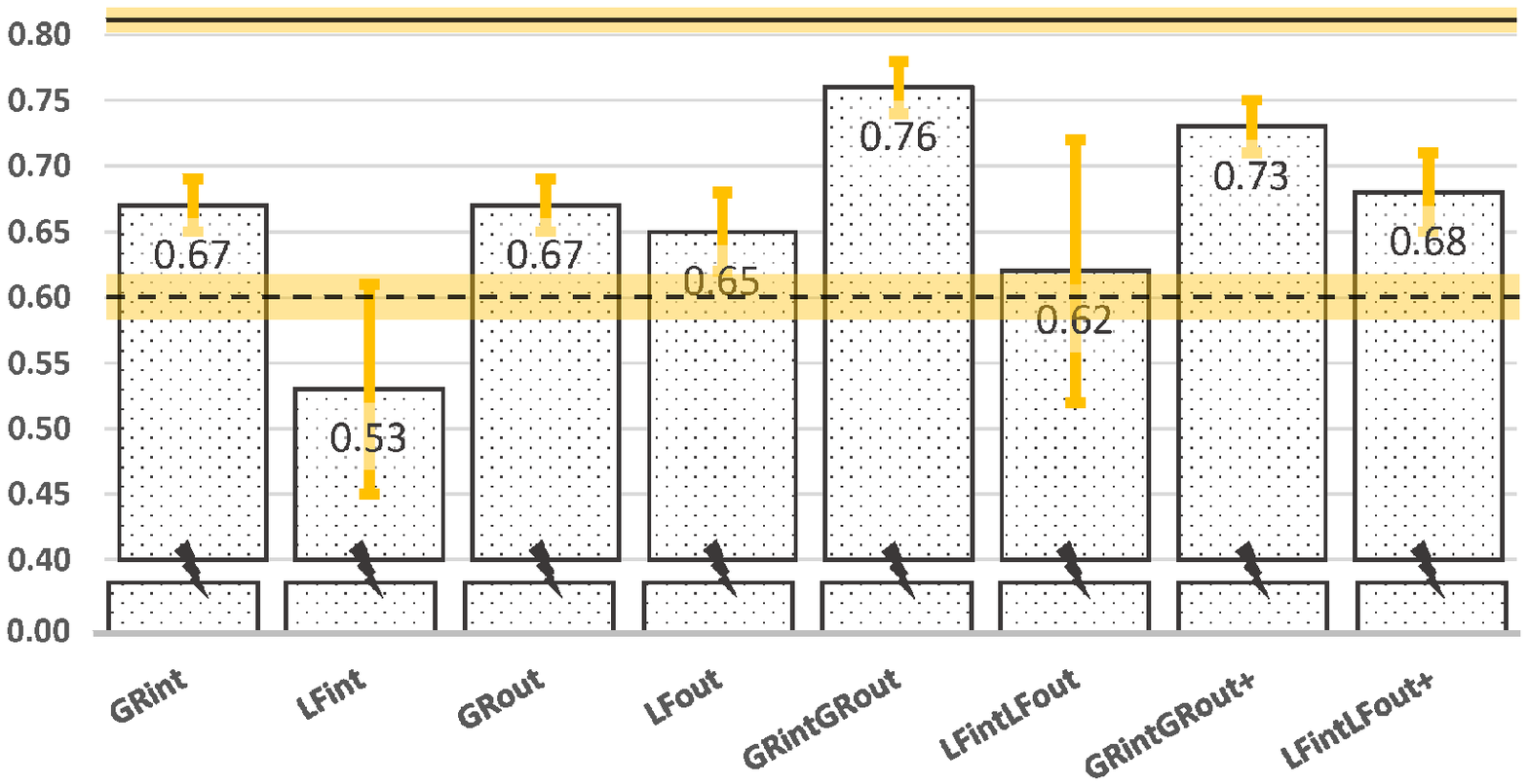}
        \label{fig:f1_test_rc_200}}%
    \qquad
    \subfloat[TestRj]{
        \includegraphics[width=7.9cm]{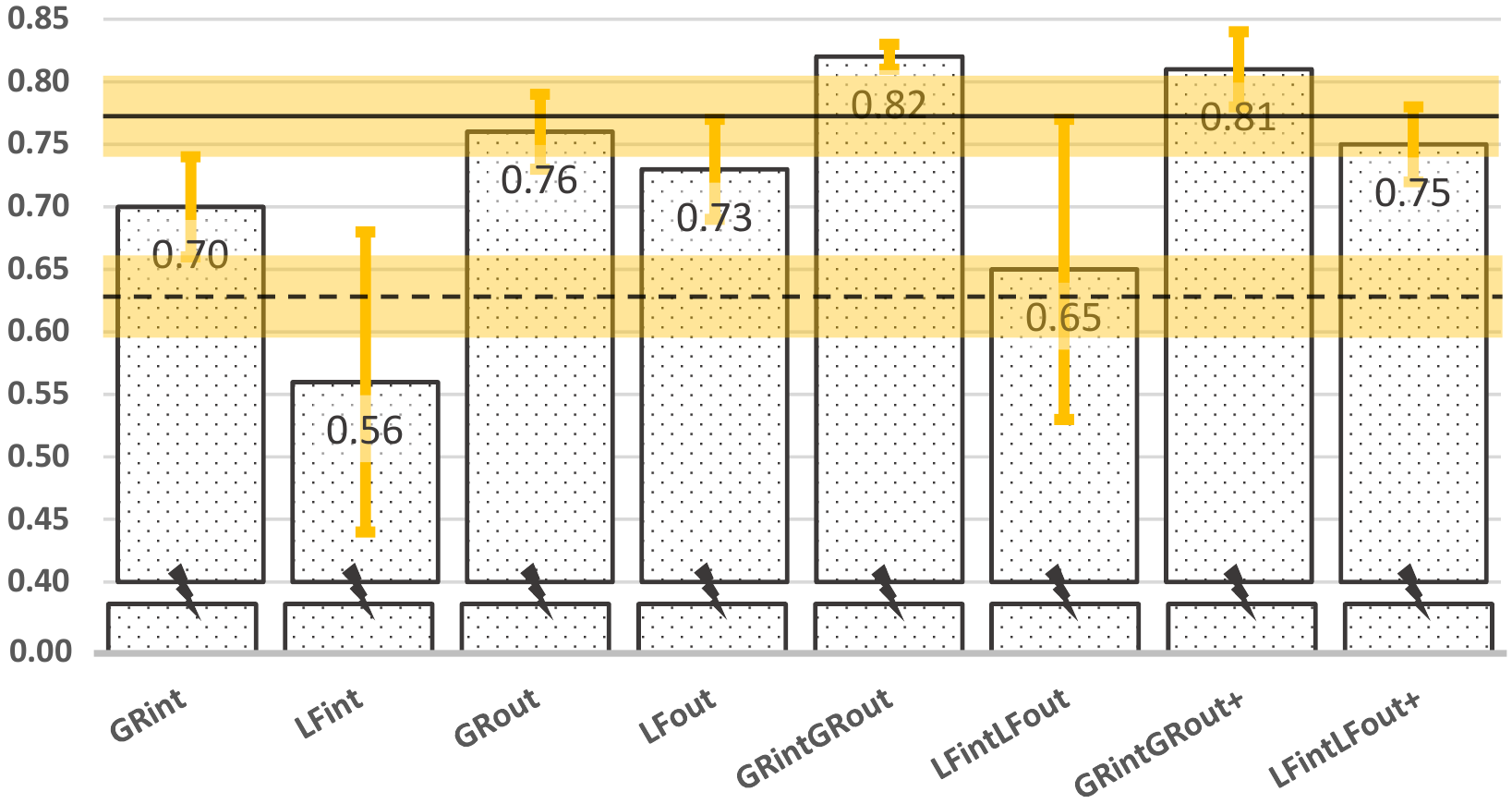}
        \label{fig:f1_test_rj_200}}%
    \qquad
    \includegraphics[width=7.5cm]{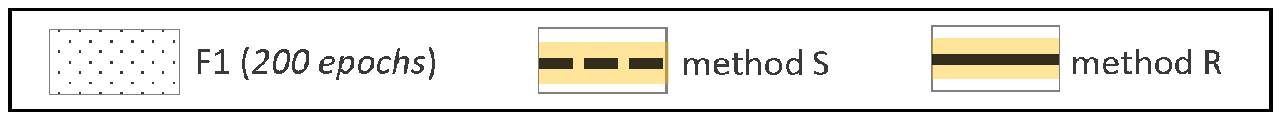}%
    \caption{F1-Score of adversarial adaptation methods compared to upper bound and lower bound methods.}%
    \label{fig:f1score}%
\end{figure}

\subsection{Experimental Protocols}

Our localization model, $M$ has the \textit{encoder-decoder} 
architecture from
\cite{conf:mmsp:lemoing2019}.
We also adopted the same input features which are 
\ac{STFT} detailed in~\cite{conf:mmsp:lemoing2019}.
These are first processed in individual encoders for each microphone array, 
merged and then decoded together.
Weights between array-encoders are shared.
Discrimination is conducted on the merged encoded features for \ac{dint}, and 
localization heatmaps for \ac{dout}.
The discriminator architecture for \ac{dint} consists of $4$ dense layers while \ac{dout} is composed of $4$ 
convolutional layers followed by $3$ dense layers.
A ReLu activation function is used for each layer, except the last one which 
has a sigmoid activation.
ADAM optimizer is used with an initial learning rate of $10^{-4}, 
\beta=(0.5, 0.999)$ and a batch size of $256$ samples.

To evaluate, we used precision, recall and F1-score as described 
in~\cite{conf:mmsp:lemoing2019}.
The source coordinates are extracted from the output heatmap as predicted 
keypoints.
These are matched with the groundtruth keypoints with a resolution threshold 
of 
$1$m.
Based on the matches, we count \ac{TP}, \ac{FP} and \ac{FN}, which are used to 
get the F1-score.
\ac{RMSE} of the matches is provided as an additional indicator of the quality.


Each method is trained for $200$ epochs and this
model is tested on both \textbf{TestRc} and 
\textbf{TestRj}.
For all domain adaptation methods, we also present results with model 
selection (from the $200$ epochs) using the best F1-score on \textbf{ValidR}.
Although this is a common measure against overfitting, it can only be done by 
using some real labeled data so we separate this result into the \textit{Sel.} 
column. 

We report the mean and standard deviation of 5 independently trained models with
different random seeds.
\begin{itemize}[wide,labelwidth=!,labelindent=0pt, noitemsep, topsep=0pt]

\item \textbf{\textit{Lower bound}} method \textbf{S} is a supervised 
method trained on labeled synthetic data (TrainS).
It represents the minimum performance without any knowledge of the target 
domain.

\item \textbf{\textit{Upper bound}} method \textbf{R} is a supervised method 
trained on labeled real-world data (TrainR).
It is an approximation of the maximum performance~\cite{jour:springer:ben-david2010}, when a 
lot of effort is spent to label real-world data which we would like to avoid in 
practice.

\item \textbf{\textit{Others}} refers to methods modifying the training data.
Method \textbf{R\&S} uses data augmentation by using TrainR and TrainS, both 
with labels, to train $M$.
It sacrifices data quality by mixing domains, but has the same data quantity used in the adversarial methods.
The main burden is that it requires labels for the real-world data.
Method \textbf{S+} uses domain randomization~\cite{jour:iros:tobin2017} by using
TrainS+ instead of TrainS in method S.

\item \textbf{\textit{Adversarial Adaptation}} methods are the variations of 
\textit{unsupervised domain adaptation using adversarial learning} that we 
described.
These are trained on TrainS with labels and TrainR without labels.
The studied methods are: 
\textbf{G\ts{r}int}, \textbf{L\ts{f}int}, 
\textbf{G\ts{r}out}, \textbf{L\ts{f}out},
and ensemble methods 
\textbf{G\ts{r}intG\ts{r}out} and 
\textbf{L\ts{f}intL\ts{f}out}.
\textit{G\ts{r}} and \textit{L\ts{f}},
refer to gradient reversal and label flipping respectively.
\textit{int} and \textit{out} are short for the discrimination level
\ac{dint} and \ac{dout}.
We also considered \textbf{G\ts{r}intG\ts{r}out+} and 
\textbf{L\ts{f}intL\ts{f}out+} which used TrainS+ instead of TrainS.

\end{itemize}

\subsection{Results and Discussion}

\input{tables/results0a_200}

Insights about the localization performance across all studied methods for TestRc can be found in~\cref{fig:f1_test_rc_200} and~\cref{tab:results0a_200}.
For method \textbf{S}, which is our lower bound baseline, there are not many 
detections (low recall) but they are mostly correct (relatively high 
precision). However, for these correct predictions, the RMSE is relatively high 
which indicates that the localization quality is low.
In comparison, method \textbf{R} which uses labeled real-world data, yields 
significantly better performance across all metrics compared to method 
\textbf{S}. 
In theory~\cite{jour:springer:ben-david2010}, our domain adaptation methods should 
have results within this range.
Method \textbf{R\&S} marginally improves performance 
compared to method \textbf{R} on TestRc.
Method \textbf{S+} shows a slight improvement over \textbf{S} but is still far 
from method \textbf{R} and \textbf{R\&S}.
Most domain adaptation methods, especially the ensemble discrimination methods, improve the performance compared to the \textit{lower bound} method \textbf{S}.
In particular, the improvement of the recall for some of the proposed methods compared to \textbf{S} is significant and leads to higher F1-score as shown in~\cref{fig:f1_test_rc_200}.

~\cref{fig:f1_test_rj_200} shows F1-score of TestRj.
The results of domain adaptation seems to be even better on TestRj even though the model is trained on classifical and funk excerpts.
We speculate that this is due to TrainS being more similar to TestRj in terms of spectral features.
This explains why lower bound \textbf{S} is slightly higher for TestRj while upper bound \textbf{R} is slightly lower.

One of our proposed ensemble discrimination methods, \textbf{G\ts{r}intG\ts{r}out}, lead to the best results across all test sets.
From \cref{tab:results0a_200} we can see that methods with \ac{dint} lead to a higher precision and a lower \ac{RMSE} while \ac{dout} mainly increases recall.
We think that aligning the intermediate features improves the existing detections while \ac{dout} encourages more detections but at a higher risk of false positives.
This interpretation is further supported by our ensembling results.
Whereas, \ac{dint} and \ac{dout} can independently lead to a similar F1-score, combining them gives the best results.
The best emsemble method is able to perform almost as good as method \textbf{R} while not having any label for the real data.
Our attempt to further improve the ensembling results with randomization (using TrainS+ instead of TrainS) did not show any clear benefit.

\input{tables/results0b}

Up to this point, the domain adaptation methods that have been presented did not use any label for the real data.
As we can see in \cref{tab:results0b}, model selection can further improve performance across all 
methods at the price of requiring a small labeled real-world dataset for validation (ValidR).
Methods based on G\ts{r} and L\ts{f} generally provide improved F1-score 
compared to \textbf{S} except for \textbf{L\ts{f}int}.
Methods based on L\ts{f} have lower performance and higher variance when using 
the last epoch model.
Model selection mitigates both issues but G\ts{r} is still 
marginally better.

Finally, we note that although most of our results are contrary to the results in~\cite{conf:icassp:He2019}, adversarial methods are notoriously difficult to train~\cite{conf:nips:lucic2014}. The high variance of L\ts{f} in our experimental results further supports this.

\section{Conclusion}
\label{sec:conclusion}

In this paper we studied adversarial domain adaptation methods from synthetic to 
real-world data for deep learning based multiple sound source 2D localization.
These methods, using labeled synthetic data and unlabeled real-world 
data, yield a significant improvement compared to solely training on synthetic 
data at a much lower cost than training on labeled real-world data.
Moreover, extensive experiments showed that our novel ensemble discrimination method
led to the best results, improving over single discrimination methods by a clear margin.

\bibliographystyle{IEEEtran}

\bibliography{reference}

\end{document}

%% file: acrodef.tex
\acrodef {BCE}[BCE]{binary cross entropy}
\acrodef {MAE}[MAE]{mean average error}
\acrodef {RMSE}[RMSE]{Root mean square error}
\acrodef {MSE}[MSE]{mean squared error}
\acrodef {NMS}[NMS]{non-maximum suppression}
\acrodef {DNN}[DNNs]{deep neural networks}
\acrodef {TDOA}[TDOA]{time difference of arrival}
\acrodef {DOA}[DOA]{direction of arrival}
\acrodef {SSL}[SSL]{Sound source localization}
\acrodef {STFT}[STFT]{short-time Fourier transform}
\acrodef {tgrid}[TG-rep]{Tight grid}
\acrodef {hmap}[HM-rep]{Heat map}
\acrodef {rgrid}[RG-rep]{Refined grid}
\acrodef {CNN}[CNNs]{Convolutional Neural Network}
\acrodef {GAN}[GAN]{Generative Adversarial Net}
\acrodef {DANN}[DANN]{Domain Adversarial Neural Networks}
\acrodef {GRL}[GRL]{Gradient Reversal Layer}
\acrodef {dint}[int-dist]{intermediate discrimination}
\acrodef {dout}[out-dist]{output discrimination}

\acrodef {FP}[FP]{false positives}
\acrodef {FN}[FN]{false negatives}
\acrodef {TP}[TP]{true positives}

%% file: tables/results0a_200.tex
\begin{table}
\footnotesize\addtolength{\tabcolsep}{-6pt}
\begin{center}
\caption{Domain Adaptation Results on Additional Metrics}
\label{tab:results0a_200} 
\begin{tabular}{|C{1cm}|C{2cm}|C{0.75cm}|C{0.75cm}|C{0.75cm}|C{0.75cm}|C{0.75cm}|C{0.75cm}|}
\cline{3-8}
\multicolumn{2}{l|}{}                                                                       & \multicolumn{6}{C{4.5cm}|}{\textbf{TestRc (classical \& funk)}}                                                                                       \\ 
\hline
\multicolumn{2}{|C{3cm}|}{\textbf{Method}}                                                       & \multicolumn{2}{C{1.5cm}|}{\textbf{RMSE $\downarrow$}} & \multicolumn{2}{C{1.5cm}|}{\textbf{Precision $\uparrow$}} & \multicolumn{2}{C{1.5cm}|}{\textbf{Recall $\uparrow$}}  \\ 
\hline
Low B.                                                               & S                    & \multicolumn{2}{C{1.5cm}|}{.51$\pm$.01}                 & \multicolumn{2}{C{1.5cm}|}{.80$\pm$.03}                    & \multicolumn{2}{C{1.5cm}|}{.48$\pm$.03}                  \\ 
\hline
Up B.                                                                & R                    & \multicolumn{2}{C{1.5cm}|}{.34$\pm$.01}                 & \multicolumn{2}{C{1.5cm}|}{.95$\pm$.03}                    & \multicolumn{2}{C{1.5cm}|}{.71$\pm$.03}                  \\ 
\hline
\multirow{2}{*}{Others}                                              & R\&S                   & \multicolumn{2}{C{1.5cm}|}{.33$\pm$.02}                 & \multicolumn{2}{C{1.5cm}|}{.96$\pm$.03}                    & \multicolumn{2}{C{1.5cm}|}{.71$\pm$.04}                  \\ 
\cline{2-8}
                                                                     & S+                   & \multicolumn{2}{C{1.5cm}|}{.44$\pm$.00}                 & \multicolumn{2}{C{1.5cm}|}{.84$\pm$.01}                    & \multicolumn{2}{C{1.5cm}|}{.57$\pm$.01}                  \\ 
\hline
\hline
\multirow{8}{*}{\begin{tabular}[c]{@{}c@{}}Adv.\\Adap.\end{tabular}} & GRint                & \multicolumn{2}{C{1.5cm}|}{.48$\pm$.03}             & \multicolumn{2}{C{1.5cm}|}{.81$\pm$.05}                  &\multicolumn{2}{C{1.5cm}|}{ .57$\pm$.04}                    \\ 
\cline{2-8}
                                                                     & LFint                &\multicolumn{2}{C{1.5cm}|}{ .50$\pm$.03}              &\multicolumn{2}{C{1.5cm}|}{ .71$\pm$.06}             &\multicolumn{2}{C{1.5cm}|}{ .44$\pm$.09}                    \\ 
\cline{2-8}
                                                                     & GRout                &\multicolumn{2}{C{1.5cm}|}{ .51$\pm$.02}              & \multicolumn{2}{C{1.5cm}|}{.68$\pm$.02 }             &\multicolumn{2}{C{1.5cm}|}{ .67$\pm$.02}          \\ 
\cline{2-8}
                                                                     & LFout                & \multicolumn{2}{C{1.5cm}|}{.51$\pm$.02}              &\multicolumn{2}{C{1.5cm}|}{ .64$\pm$.04}                  &\multicolumn{2}{C{1.5cm}|}{ .67$\pm$.02}                    \\ 
\cline{2-8}
                                                                     & GRintGRout           & \multicolumn{2}{C{1.5cm}|}{.47$\pm$.05}              &\multicolumn{2}{C{1.5cm}|}{\textbf{.82$\pm$.05}}    &\multicolumn{2}{C{1.5cm}|}{\textbf{.72$\pm$.03}}                     \\ 
\cline{2-8}
                                                                     & LFintLFout           & \multicolumn{2}{C{1.5cm}|}{.53$\pm$.03}              &\multicolumn{2}{C{1.5cm}|}{ .65$\pm$.10}             &\multicolumn{2}{C{1.5cm}|}{ .60$\pm$.09}          \\ 
\cline{2-8}
                                                                     & GRintGRout+          &\multicolumn{2}{C{1.5cm}|}{\textbf{.46$\pm$.02}}              &\multicolumn{2}{C{1.5cm}|}{ .76$\pm$.07}              & \multicolumn{2}{C{1.5cm}|}{.71$\pm$.05}         \\ 
\cline{2-8}
                                                                     & LFintLFout+          &\multicolumn{2}{C{1.5cm}|}{ .50$\pm$.02}              & \multicolumn{2}{C{1.5cm}|}{.72$\pm$.09}              &\multicolumn{2}{C{1.5cm}|}{ .65$\pm$.02}          \\ 
\hline
\multicolumn{1}{l}{}                                                 & \multicolumn{1}{l}{} & \multicolumn{1}{l}{} & \multicolumn{1}{l}{}  & \multicolumn{1}{l}{} & \multicolumn{1}{l}{}     & \multicolumn{2}{l}{}                         
\end{tabular}

\begin{scriptsize}
\begin{center}
\vspace{-2.4em}
Results for the 200\textsuperscript{th}  epoch of training;
\textit{Low B.}, \textit{Up B.} and \textit{Adv. Adap.} refer to\\ lower bound, upper bound and adversarial adaptation respectively;\\$\uparrow$/$\downarrow$ represent the higher/lower is better respectively.
\end{center}
\end{scriptsize}

\end{center}
\normalsize
\end{table}

%% file: tables/results0b.tex
\begin{table}
\footnotesize\addtolength{\tabcolsep}{-2pt}
\begin{center}
\caption{Comparison of F1-Score with/without Model Selection.}
\label{tab:results0b} 

\begin{tabular}{|c|c|c|c|c|c|c|}
\cline{3-7}
\multicolumn{2}{l|}{}                                                              & \multicolumn{5}{c|}{ \textbf{F1-Score $\uparrow$}}                                                                                                                                                               \\ 
\hline
\multicolumn{2}{|c|}{\textbf{\textit{Test Data}}}                                  & \multicolumn{2}{c|}{\begin{tabular}[c]{@{}c@{}}\textbf{TestRc}\\\textbf{(classical\&funk)} \end{tabular}} &  & \multicolumn{2}{c|}{\begin{tabular}[c]{@{}c@{}}\textbf{TestRj}\\\textbf{(jazz)}\end{tabular}}  \\ 
\cline{1-4}\cline{6-7}
\multicolumn{2}{|c|}{\textit{\textbf{Stopping Epoch}}}                             & \textit{200th}   & \textit{Sel.}                                                                          &  & \textit{200th}   & \textit{Sel.}                                                               \\ 
\cline{1-4}\cline{6-7}
\multirow{8}{*}{\begin{tabular}[c]{@{}c@{}}Adv.\\Adap.\end{tabular}} & GRint       & .67$\pm$.02          & .71$\pm$.02                                                                                &  & .70$\pm$.04          & .77$\pm$.03                                                                     \\ 
\cline{2-4}\cline{6-7}
                                                                     & LFint       & .53$\pm$.08          & .64$\pm$.03                                                                                &  & .56$\pm$.12          & .69$\pm$.03                                                                     \\ 
\cline{2-4}\cline{6-7}
                                                                     & GRout       & .67$\pm$.02          & .72$\pm$.03                                                                                &  & .76$\pm$.03          & .79$\pm$.03                                                                     \\ 
\cline{2-4}\cline{6-7}
                                                                     & LFout       & .65$\pm$.03          & .73$\pm$.02                                                                                &  & .73$\pm$.04          & .80$\pm$.01                                                                     \\ 
\cline{2-4}\cline{6-7}
                                                                     & GRintGRout  & \textbf{.76$\pm$.02} & \textbf{.79$\pm$.01}                                                                       &  & \textbf{.82$\pm$.01} & .84$\pm$.01                                                                     \\ 
\cline{2-4}\cline{6-7}
                                                                     & LFintLFout  & .62$\pm$.10          & .74$\pm$.03                                                                                &  & .65$\pm$.12          & .81$\pm$.03                                                                     \\ 
\cline{2-4}\cline{6-7}
                                                                     & GRintGRout+ & .73$\pm$.02          & .78$\pm$.02                                                                                &  & .81$\pm$.03          & \textbf{.85$\pm$.03}                                                            \\ 
\cline{2-4}\cline{6-7}
                                                                     & LFintLFout+ & .68$\pm$.03          & \textbf{.79$\pm$.02}                                                                       &  & .75$\pm$.03          & .84$\pm$.01                                                                     \\
\hline
\end{tabular}

\begin{scriptsize}
\begin{center}
\vspace{-1.4em}
\textit{Adv. Adap.}, and \textit{Sel.} is short for adversarial adaptation, and model selection; \\$\uparrow$ represent the higher is better.
\end{center}
\end{scriptsize}

\end{center}
\normalsize
\end{table}